# MODELS OF STRANGE STARS WITH A CRUST AND STRANGE DWARFS

Yu. L. Vartanyan, A. K. Grigoryan, and T. R. Sargsyan

*Strange quark stars with a crust and strange dwarfs consisting of a compact strange quark core and an extended crust are investigated in terms of a bag model. The crust, which consists of atomic nuclei and degenerate electrons, has a limiting density of $\rho_{cr} = \rho_{drip} = 4.3 \cdot 10^{11}$ g/cm³. A series of configurations are calculated for two sets of bag model parameters and three different values of $\rho_{cr}$ ($10^9$ g/cm³ $\leq \rho_{cr} \leq \rho_{drip}$) to find the dependence of a star's mass M and radius R on the central density. Sequences of stars ranging from compact strange stars to extended strange dwarfs are constructed out of strange quark matter with a crust. The effect of the bag model parameters and limiting crust density $\rho_{cr}$ on the parameters of the strange stars and strange dwarfs is examined. The strange dwarfs are compared with ordinary white dwarfs and observational differences between the two are pointed out.*

## 1. Introduction

The hypothesis that strange quark matter, consisting of roughly equal amounts of *u, d*, and *s*-quarks with a small additive of electrons or positrons to ensure electrical neutrality may be the absolutely stable state of cold matter was first proposed by Witten [1]. Later Farhi and Jaffe [2] used an MIT bag model [3] to study the dependence of the stability of strange quark matter on the insufficiently accurately known phenomenological model parameters, the bag constant *B*, the quark-gluon interaction constant $\alpha_c$, and the strange quark mass $m_s$. It was shown that certain sets of these parameters could yield self-confining strange stars. The main properties of the latter have been examined in Refs. 4 and 5. In Refs. 6 and 7 the parameters of strange stars are compared with observational data and the problem of the parallel existence of strange and neutron stars examined.

If a variant of strange quark matter can be realized in which the excess electrical charge of the quarks is neutralized by electrons, then the latter, bound only by a Coulomb force, can partially leave the quark surface and propagate hundreds

---

of Fermis. For this reason, a thin charged layer develops at the surface of strange quark stars which resembles a capacitor where the field strength attains $10^{17}$-$10^{18}$ V/cm [4].

Since the electric field at the surface of a strange quark star is directed outward, it can sustain a crust consisting of atomic nuclei and degenerate electrons. The crust is not in chemical equilibrium with the strange quark matter and is coupled to the quark core only by gravity. The probability of tunneling transitions by the atomic nuclei is so small that the two phases can coexist for an essentially infinite time [4]. Since uncharged free neutrons can pass unhindered through the electrostatic barrier and be absorbed by the strange quark matter, the maximum density of the crust must be limited by the density at which neutrons escape from nuclei (neutron drip density), $\rho_{drip} = 4.3 \cdot 10^{11}$ g/cm$^3$. A strange star may acquire a crust when it is formed or through accretion of material. The formation and structure of crusts in strange stars has been studied in Refs. 8 and 9.

For strange stars with masses $M > 0.5 M_\odot$, the thickness and mass of the crust are negligible compared to the star's radius and mass. The situation differs for strange stars with low masses. If the mass of a strange star $M < 0.02 M_\odot$, then the shell swells significantly and its maximum radius is on the order of that of white dwarfs. Unlike ordinary white dwarfs, these configurations, which are referred to as strange dwarfs, have a core consisting of a strange star with a small size and mass. Note that a second kind of strange dwarf can also exist, in which the central quark core cannot exist independently in the form of a strange star, but is in thermodynamic equilibrium with a density jump with a shell that contains degenerate neutrons as well as atomic nuclei and degenerate electrons [10,11]. Configurations of this sort are not considered here.

In this paper we study models of strange stars with a crust and strange dwarfs. Calculations are done for two sets of bag model parameters, on which the parameters of the strange quark core depend, and three values for the limiting density of the crust. In ordinary white dwarfs the central density cannot exceed $10^9$ g/cm$^3$ because of neutronization of the atomic nuclei (the configuration becomes unstable). In strange dwarfs, on the other hand, densities more than two orders of magnitude higher can exist; there the high Fermi momentum level of the electrons ($p_F(e)/m_e c \leq 45$) makes it possible for anomalous atomic nuclei, highly overloaded with neutrons ($A \sim 120$, $A/Z \sim 3$), to exist. The stability of these models is ensured by the presence of a small quark core [12,13].

Here we introduce the integral parameters of strange stars and strange dwarfs. Strange dwarfs are compared with their non-strange analogs, ordinary white dwarfs, and observational differences between the two are pointed out.

## 2. Equation of state

Neglecting the gap of several hundred Fermis between the strange quark matter and the crust, we use an equation of state consisting of two parts which are coupled by a pressure continuity condition. The first part describes the normal matter in the Ae-phase. We have used tabulated data on the Baym-Pethick-Sutherland equation of state [14] matched to the Feynman-Metropolis-Teller equation of state [15] at a density of $\rho = 10^4$ g/cm$^3$.

The second part corresponds to the strange quark matter, for which we use an MIT bag model. When a crust is present, the pressure at the boundary of the quark core does not go to zero, but corresponds to the transition pressure $P_{tr}$. The density dependence of the pressure for strange stars with a core is illustrated schematically in Fig. 1 of Ref. 16. We have considered the equations of state for two sets of bag model parameters with the parameters listed in Table 1.

TABLE 1. Parameters of the Equation of State of Strange Quark Matter

|         | $B$ (MeV/fm$^3$) | $m_s$ (MeV) | $\alpha_c$ | $\bar{\varepsilon}_b$ (MeV) | $n_{min} = n_s$ (fm$^{-3}$) |
|---------|------------------|-------------|------------|-----------------------------|------------------------------|
| Model 1 | 50               | 175         | 0.05       | -64.9                       | 0.257                        |
| Model 2 | 60               | 175         | 0.05       | -28.6                       | 0.296                        |

Here, as in Refs. 16 and 17, we use an expanded form of the equation of state. For these sets the average energy per baryon has a negative minimum that depends on the baryon concentration, which ensures that the strange quark matter is bound. These quantities, which characterize the surface of the quark core, are also listed in the last two columns of Table 1.

Attempts have been made [18,19] to review the maximum allowable limiting density of the crust. In the first paper, the effect of the chemical potential of the electrons at the surface of a strange star was considered and in the second, the mechanical equilibrium of the crust. The configuration with $\rho_{cr} = \rho_{drip}$ is of greatest physical interest, but any other configurations with lower values of $\rho_{cr}$ can be realized; thus, in order to study the effect of the limiting crust density on the integral parameters of strange stars, we use a range of $\rho_{cr}$ extending from the maximum densities in white dwarfs, $10^9$ g/cm$^3$, to $\rho_{drip} = 4.3 \cdot 10^{11}$ g/cm$^3$.

## 3. Results of the calculations

The relativistic equations of stellar equilibrium (the Tolman-Oppenheimer-Volkov equations) [20] were integrated to find the main parameters of spherically symmetric superdense stars. The star's radius $R$ and total mass $M$, as well as the mass $M_{core}$ and radius $R_{core}$ of the quark core were calculated for a series of configurations depending on the central density $\rho_c$.

Tables 2 and 3 list the calculated results for models 1 and 2 for three values of $\rho_{cr}$. The calculations encompass the entire range of realizable central densities rc corresponding to configurations in the range from massive strange stars to strange dwarfs. The labels a and b denote the configurations with the maximum and minimum masses of strange stars with a crust, respectively, and c, the configuration with the maximum masses for strange dwarfs. The sequence of strange dwarfs terminates at the configuration labeled d, for which the central density is no longer sufficient for existence of a quark core.

Let us begin by studying the configuration with the maximum crust density $\rho_{cr} = \rho_{drip} = 4.3 \cdot 10^{11}$ g/cm$^3$. The total mass $M$ of the star is plotted as a function of the central density $\rho_c$ in Fig. 1. Two curves, corresponding to models 1 and 2, are plotted for comparison. The more rigid equation of state (model 1) evidently leads to a leftward shift of the $M(\rho_c)$ curve in the figure. This happens because the more rigid equation of state has a higher pressure for a given density of the material, which leads to greater maximum mass and radius of the star for a lower central density.

The total mass $M$ is plotted in Fig. 2 as a function of radius $R$ in models 1 and 2 for $\rho_{cr} = 4.3 \cdot 10^{11}$ g/cm$^3$. Massive strange stars with the maximum central density for the quark core lie to the left in the figure. The maximum mass of the strange stars with a crust is $1.95 M_\odot$ for model 1 and $1.79 M_\odot$ for model 2. The extent of the crust is minimal for these configurations. For strange stars with masses $M \approx 1.1 \div 1.8 M_\odot$, which are typical of observed superdense stars, the crust thickness is on the order of 200-500 m.

TABLE 2. Basic Parameters of the Sequence of Strange Stars with a Crust and Strange Dwarfs for Model 1

| $\rho_{cr}$ (g/cm³) | $\rho_c$ ($10^{14}$ (g/cm³)) | $M_{core}$, $M_\odot$ | $M$, $M_\odot$ | $R_{core}$, km | $R$, km |
|---|---|---|---|---|---|
| 1 | 2 | 3 | 4 | 5 | 6 |
| $4.3 \cdot 10^{11}$ | 3.96059 (d) | 0 | 0.6762 | 0 | 484.1 |
| | 3.98832 | 0.00509 | 0.7966 | 1.828 | 816.3 |
| | 4.00047 | 0.00870 | 0.8510 | 2.184 | 1165.2 |
| | 4.01591 (c) | 0.01405 | 0.9646 | 2.561 | 2347.0 |
| | 4.02448 | 0.01732 | 0.7232 | 2.745 | 5280.7 |
| | 4.02681 | 0.01824 | 0.2038 | 2.793 | 9960.2 |
| | 4.02699 | 0.01831 | 0.0972 | 2.797 | 10823.4 |
| | 4.02705 | 0.01834 | 0.0577 | 2.798 | 10415.3 |
| | 4.02713 | 0.01837 | 0.0234 | 2.800 | 5172.2 |
| | 4.02722 | 0.01841 | 0.0193 | 2.801 | 1710.2 |
| | 4.02755 (b) | 0.01854 | 0.0188 | 2.808 | 452.6 |
| | 4.03394 | 0.02115 | 0.0212 | 2.934 | 32.3 |
| | 4.11837 | 0.06263 | 0.0627 | 4.201 | 7.4 |
| | 4.49129 | 0.29948 | 0.2995 | 6.997 | 8.1 |
| | 6.10283 | 1.09477 | 1.0948 | 10.354 | 10.8 |
| | 20.1318 (a) | 1.94517 | 1.9452 | 10.857 | 11.1 |
| | 32.0556 | 1.89849 | 1.8985 | 10.263 | 10.4 |
| $10^{10}$ | 3.96059 (d) | 0 | 0.9108 | 0 | 1345.6 |
| | 3.97180 (c) | 0.00133 | 1.0145 | 1.170 | 2289.9 |
| | 3.98009 | 0.00303 | 0.7938 | 1.538 | 5136.1 |
| | 3.98246 | 0.00359 | 0.3004 | 1.627 | 9991.3 |
| | 3.98285 | 0.00364 | 0.0227 | 1.640 | 18232.6 |
| | 3.98286 | 0.00365 | 0.0135 | 1.641 | 16837.4 |
| | 3.98287 | 0.00366 | 0.0043 | 1.642 | 6028.0 |
| | 3.98289 | 0.00367 | 0.0038 | 1.643 | 1680.8 |
| | 3.98294 (b) | 0.00369 | 0.0037 | 1.645 | 534.6 |
| | 3.98304 | 0.00373 | 0.0038 | 1.648 | 225.3 |
| | 3.98388 | 0.00388 | 0.0039 | 1.678 | 39.1 |
| | 4.03156 | 0.02018 | 0.0202 | 2.888 | 4.2 |
| | 4.19693 | 0.10848 | 0.1085 | 5.032 | 5.6 |
| | 5.42628 | 0.82808 | 0.8281 | 9.579 | 9.8 |
| | 6.87139 | 1.31259 | 1.3126 | 10.834 | 11.0 |
| | 20.1319 (a) | 1.94518 | 1.9453 | 10.858 | 10.9 |
| | 32.0558 | 1.89859 | 1.8986 | 10.264 | 10.3 |

TABLE 2. (continued)

| 1 | 2 | 3 | 4 | 5 | 6 |
|---|---|---|---|---|---|
| | 3.96059 (d) | 0 | 1.0192 | 0 | 2338.9 |
| | 3.96115 | 0.00001 | 1.0184 | 0.269 | 2430.9 |
| | 3.96884 | 0.00080 | 0.7625 | 1.005 | 5519.5 |
| | 3.97081 | 0.00113 | 0.3086 | 1.117 | 10057.8 |
| | 3.97118 | 0.00115 | 0.0530 | 1.137 | 16927.0 |
| | 3.97121 | 0.00116 | 0.0247 | 1.138 | 21747.8 |
| | 3.97122 | 0.00117 | 0.0023 | 1.139 | 12267.0 |
| | 3.97123 | 0.00118 | 0.0013 | 1.140 | 1916.9 |
| $10^9$ | 3.97125 (b) | 0.00119 | 0.0012 | 1.141 | 565.3 |
| | 3.97147 | 0.00127 | 0.0013 | 1.152 | 50.7 |
| | 3.97280 | 0.00148 | 0.0015 | 1.220 | 9.5 |
| | 3.99548 | 0.00716 | 0.0072 | 2.047 | 3.0 |
| | 4.31481 | 0.18337 | 0.1834 | 5.973 | 6.2 |
| | 5.48441 | 0.85445 | 0.8545 | 9.666 | 9.8 |
| | 9.84955 | 1.72758 | 1.7276 | 11.348 | 11.4 |
| | 20.1322 (a) | 1.94558 | 1.9456 | 10.859 | 10.9 |
| | 32.0559 | 1.89869 | 1.8987 | 10.265 | 10.3 |

As the central density of the core is reduced, the mass and radius of the configuration begin to decrease, while the extent of the crust gradually increases. For the maximum value $\rho_{cr}$ the minimum radius of strange stars with a crust is on the order of $R_{min} \approx 6.5 \div 7.5$ km for a mass $M \approx 0.06 \div 0.09 M_\odot$. The situation changes for strange stars with a crust that have lower masses; there a sharp rise in the thickness of the crust is observed, so that the star's radius increases. The star's behavior resembles that of a neutron star, for which a sharp rise in the radius is also observed for low masses owing to swelling of the shell. Note that for bare strange stars without an outer shell, the radius increases with rising mass over almost the entire curve and only at the very maximum is this dependence the same as for stable neutron stars.

At a certain central density there is a minimum in the mass, where $dM/d\rho_c = 0$ (points b in Figs. 1 and 2). For the models considered here the minimum mass of the strange stars with a crust is on the order of $M_{min} \approx 0.017 \div 0.019 M_\odot$ with a radius $R \approx 450$ km. The bulk of the mass in these configurations is, as before, concentrated in the quark core.

With further reductions in the central density, the mass of the configuration gradually begins to rise owing to the mass of the crust; here the radius continues to increase rapidly because of the increasing crust thickness. This is the region of strange dwarfs. According to the calculations of Glendenning, et al. [12,13], who have studied the stability of strange dwarfs using the standard Chandrasekhar method [21,22], configurations with $dM/d\rho_c < 0$, as opposed to the case of ordinary white dwarfs and neutron stars, are stable with respect to radial fluctuations. We have not examined questions of stability separately in our work.

TABLE 3. Basic Parameters of the Sequence of Strange Stars with a Crust and Strange Dwarfs for Model 2

| $\rho_{cr}$, g/cm$^3$ | $\rho_{cr}$, $10^{14}$ g/cm$^3$ | $M_{core}$, $M_\odot$ | $M$, $M_\odot$ | $R_{core}$, km | $R$, km |
|---|---|---|---|---|---|
| 1 | 2 | 3 | 4 | 5 | 6 |
| | 4.73524 (d) | 0 | 0.6762 | 0 | 484.1 |
| | 4.76265 | 0.00355 | 0.7757 | 1.528 | 729.5 |
| | 4.80176 (c) | 0.01310 | 0.9700 | 2.358 | 2449.0 |
| | 4.81037 | 0.01564 | 0.7794 | 2.500 | 4812.5 |
| | 4.81349 | 0.01659 | 0.3332 | 2.550 | 8825.8 |
| | 4.81406 | 0.01677 | 0.0739 | 2.559 | 11104.9 |
| | 4.81424 | 0.01683 | 0.0183 | 2.562 | 2669.0 |
| $4.3 \cdot 10^{11}$ | 4.81442 | 0.01688 | 0.0172 | 2.564 | 868.9 |
| | 4.81459 (b) | 0.01693 | 0.0171 | 2.567 | 448.7 |
| | 4.81600 | 0.01737 | 0.0174 | 2.589 | 118.4 |
| | 4.82162 | 0.01915 | 0.0192 | 2.674 | 31.7 |
| | 4.89166 | 0.04464 | 0.0447 | 3.539 | 7.2 |
| | 4.98925 | 0.08698 | 0.0870 | 4.409 | 6.5 |
| | 5.78574 | 0.47890 | 0.4791 | 7.634 | 8.4 |
| | 8.01201 | 1.17039 | 1.1704 | 9.856 | 10.2 |
| | 23.9710 (a) | 1.78609 | 1.7861 | 9.951 | 10.1 |
| | 32.8297 | 1.76499 | 1.7650 | 9.588 | 9.8 |
| | 4.73524 (d) | 0 | 0.9062 | 0 | 1314.7 |
| | 4.74875 (c) | 0.00124 | 1.0150 | 1.076 | 2309.3 |
| | 4.75814 | 0.00272 | 0.8144 | 1.398 | 4942.9 |
| | 4.76107 | 0.00326 | 0.3543 | 1.484 | 9350.4 |
| | 4.76158 | 0.00335 | 0.1018 | 1.499 | 13667.5 |
| | 4.76168 | 0.00336 | 0.0232 | 1.501 | 18678.6 |
| | 4.76170 | 0.00337 | 0.0038 | 1.502 | 5545.9 |
| | 4.76175 | 0.00338 | 0.0034 | 1.503 | 871.5 |
| $10^{10}$ | 4.76178 (b) | 0.00339 | 0.0034 | 1.504 | 524.7 |
| | 4.76248 | 0.00349 | 0.0035 | 1.524 | 53.6 |
| | 4.77046 | 0.00516 | 0.0052 | 1.729 | 7.0 |
| | 4.85257 | 0.02967 | 0.0297 | 3.094 | 4.0 |
| | 5.55328 | 0.36858 | 0.3686 | 7.033 | 7.3 |
| | 7.20015 | 0.98378 | 0.9838 | 9.431 | 9.6 |
| | 10.0511 | 1.45708 | 1.4571 | 10.303 | 10.4 |
| | 23.9711 (a) | 1.78619 | 1.7862 | 9.952 | 10.0 |
| | 32.8298 | 1.76509 | 1.7651 | 9.589 | 9.6 |

| 1 | 2 | 3 | 4 | 5 | 6 |
|---|---|---|---|---|---|
| | 4.73524 (d) | 0 | 1.0193 | 0 | 2334.3 |
| | 4.74023 | 0.00020 | 0.9958 | 0.656 | 3229.1 |
| | 4.74346 | 0.00050 | 0.8785 | 0.841 | 4411.1 |
| | 4.74610 | 0.00080 | 0.6460 | 0.966 | 6735.3 |
| | 4.74769 | 0.00100 | 0.1615 | 1.034 | 12429.1 |
| | 4.74788 | 0.00101 | 0.0254 | 1.040 | 21887.8 |
| | 4.74789 | 0.00102 | 0.0043 | 1.041 | 16990.1 |
| | 4.74790 | 0.00108 | 0.0012 | 1.042 | 3355.0 |
| $10^9$ | 4.74791 (b) | 0.00109 | 0.0011 | 1.043 | 553.7 |
| | 4.74809 | 0.00119 | 0.0012 | 1.050 | 66.4 |
| | 4.75022 | 0.00148 | 0.0015 | 1.133 | 7.3 |
| | 4.76987 | 0.00498 | 0.0050 | 1.715 | 2.7 |
| | 5.17438 | 0.17747 | 0.1775 | 5.564 | 5.7 |
| | 6.33598 | 0.70858 | 0.7086 | 8.595 | 8.7 |
| | 10.6425 | 1.51009 | 1.5101 | 10.353 | 10.4 |
| | 23.9712 (a) | 1.78629 | 1.7863 | 9.953 | 10.0 |
| | 32.8299 | 1.76518 | 1.7652 | 9.590 | 9.6 |

The radii of strange dwarfs in configurations with lower central densities of the core reach $R_{max} \approx 10800 \div 11100$ km for masses of $M \approx 0.07 \div 0.1 M_\odot$. Here, also, the mass of the shell is considerably greater than that of the quark core, but is still far from its maximum value in terms of its order of magnitude.

Then the rise in mass becomes more rapid because of the increasing mass of the crust as the configuration radius decreases. Strange dwarfs lose stability with respect to radial fluctuations at the point c [12,13]. This point corresponds to the configurations of maximally heavy strange dwarfs. Here the mass of the star is on the order of $M \approx 0.96 \div 0.97 M_\odot$ for a radius $R \approx 2350 \div 2450$ km. It is noteworthy that the central density of the quark core does not change by more than 0.5% along the entire path from point b to point c.

We, therefore, obtain a sequence of stable stars made of strange quark matter with a crust ranging from compact strange stars to extended strange dwarfs. The strange dwarfs are stable exclusively because of the compact quark core [12,13], without which they would lie an unstable region between white dwarfs and neutron stars.

As the limiting crust density decreases, changes occur in the integral parameters of the strange stars and strange dwarfs. Figure 3 is a plot of the total mass *M* as a function of radius *R* for two different values of the limiting crust density $\rho_{cr}$ in the case of model 1. Whereas the maximum mass of the strange stars with a crust is essentially independent of $\rho_{cr}$, the minimum mass of the configuration is very sensitive to the magnitude of this parameter. When $\rho_{cr}$ is reduced from $4.3 \cdot 10^{11}$ to $10^9$ g/cm$^3$, the minimum mass of the strange stars changes by more than an order of magnitude, reaching $M_{min} \approx 0.0011 \div 0.0012 \, M_\odot$. Here the radius of the core falls from $R_{core} \approx 2.6 \div 2.8$ km to 1.04-1.14 km, while the radius of the configuration exceeds 550 km.

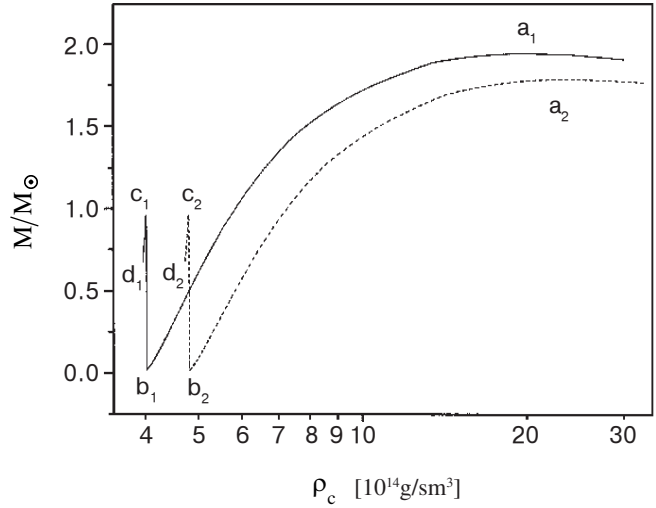

Fig. 1. The total mass $M$ as a function of the central density $\rho_c$ for configurations with a limiting crust density $\rho_{cr} = 4.3 \cdot 10^{11}$ g/cm$^3$ for models 1 (smooth curve) and 2 (dashed curve). The labels $a_1$, $a_2$ and $b_1$, $b_2$, respectively, denote configurations with maximum and minimum masses for strange stars, $c_1$, $c_2$, configurations with maximum mass for strange dwarfs, and $d_1$, $d_2$, configurations without a central quark core.

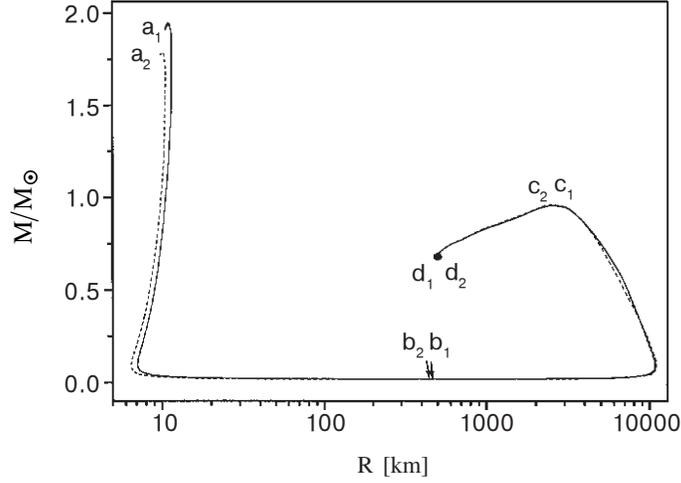

Fig. 2. Total mass $M$ as a function of radius $R$ for configurations with the limiting density $\rho_{cr} = 4.3 \cdot 10^{11}$ g/cm$^3$ for model 1 (smooth curve) and 2 (dashed curve). The labels a, b, c, and d are as in Fig. 1.

Our results differ significantly from the case of neutron stars, for which the minimum mass reaches $M \approx 0.1 M_\odot$ with a radius of 200 km [23]. The minimum mass of strange stars for this range of limiting densities is an order of magnitude or two smaller than for neutron stars.

For strange stars and strange dwarfs with a lower limiting density $\rho_{cr}$ of the crust, configurations with a wider range of values for the stellar radius are realizable. Thus, for $\rho_{cr} = 10^9$ g/cm$^3$ the minimum radius of the strange stars with a crust is on the order of $R_{min} \approx 3$ km, while the maximum radius for the strange dwarfs is $R_{max} \approx 22000$ km. Note that the choice of model for the strange quark matter essentially has no effect on $R_{min}$ and $R_{max}$.

In order to compare the strange dwarfs with ordinary white dwarfs, we show three curves in Fig. 4: curves 1 and 2 are the configurations of strange dwarfs with different values of the limiting crust density: $\rho_{cr} = 4.3 \cdot 10^{11}$ g/cm$^3$ and $\rho_{cr} = 10^9$ g/cm$^3$. Curve 3 corresponds to a sequence of ordinary white dwarfs constructed from the data of Ref. 14. The mass of the maximally heavy white dwarf is $M_{wd\,max} = 1.02 M_\odot$ with a radius $R \approx 2400$ km and a central density $\rho_c \approx 10^9$ g/cm$^3$.

The plots of the mass as a function of radius for strange dwarfs with $\rho_{cr} = 10^9$ g/cm$^3$ and for the ordinary white dwarfs are essentially identical, although the directions of variation in the central density are opposite (given the position of the configurations in the figure). Thus, for the stable branch of the ordinary white dwarfs, there is a characteristic rise in mass as the central density of the configuration is raised, while the mass of the strange dwarfs increases as the central density of the quark core is lowered. The characteristic difference in the directions of the rise in mass in white and strange dwarfs as a function of the central density is evident from the data in Table 4.

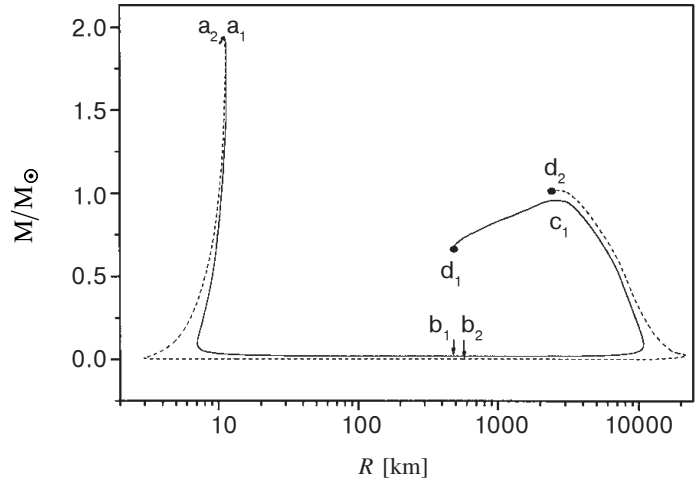

Fig. 3. Total mass $M$ as a function of radius $R$ for configurations with a limiting crust density $\rho_{cr} = 4.3 \cdot 10^{11}$ g/cm$^3$ (smooth curve, labels $a_1$, $b_1$, $c_1$, and $d_1$) and for $\rho_{cr} = 10^9$ g/cm$^3$ (dashed curve, labels $a_2$, $b_2$, and $d_2$) for model 1. The labels a, b, c, and d are as in Fig. 1.

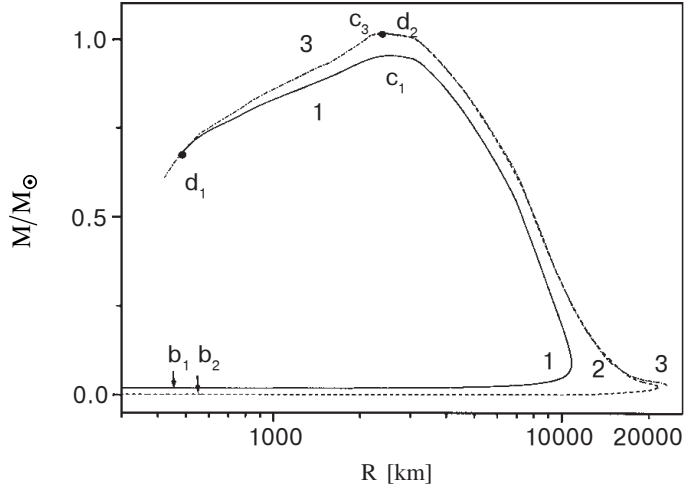

Fig. 4. Total mass $M$ as a function of radius $R$ for strange dwarfs (1, $\rho_{cr} = 4.3\cdot 10^{11}$ g/cm$^3$, smooth curve, labels $b_1$, $c_1$, and $d_1$; 2, $\rho_{cr}=10^9$ g/cm$^3$, dashed curve, labels $b_2$ and $d_2$) for model 1 and white dwarfs (3, dashed curve dot-dashed). The labels b, c, and d are as in Fig. 1. The label $c_3$ denotes the configuration with maximum mass for white dwarfs.

Among the strange dwarfs, the mass and radius of the quark core decrease in parallel with a reduction in the central density. At the point $d_2$ of Fig. 4 the central density is no longer sufficient for the existence of a strange quark core. Note that this holds until the maximum mass configuration is reached. Thus, in this case at the point $d_2$, where the strange quark core ceases to exist, the configurations of strange dwarfs transform smoothly into a branch of ordinary stable white dwarfs without loss of stability.

A similar smooth transition fom strange to ordinary white dwarfs does not occur for configurations with a limiting crust density $\rho_{cr}=10^9$ g/cm$^3$, since the existence of these configurations is only possible as the result of the presence of a quark core which, lying at the center of the star, stabilizes it. Strange dwarfs of this sort represent a qualitatively new class of superdense heavenly objects. They can fill the existing gap in the Hertzsprung-Russell diagram between stable neutron stars and ordinary white dwarfs.

We now attempt to find the differences between strange dwarfs with maximally high limiting crust densities $\rho_{cr} = \rho_{drip}$ and ordinary white dwarfs. For comparison the parameters of ordinary white dwarfs and strange dwarfs with the same masses in the interval from $M = 0.02\,M_\odot$ to $0.96\,M_\odot$ are listed in Table 4. Note that for the ordinary white dwarfs with low masses ($M \leq 0.03\,M_\odot$), the integral parameters have been calculated using only the equations of state [15]. The radii for these models differ substantially from their strange analogs, which are much more compact than ordinary white dwarfs and, therefore, have luminosities that are lower by more than an order of magnitude for a given surface temperature.

TABLE 4. Parameters of White and Strange Dwarfs (Model 1) with the Same Masses

| | White dwarfs | | Strange dwarfs | | | |
|---|---|---|---|---|---|---|
| $M$, $M_\odot$ | $\rho_c$, g/cm$^3$ | $R$, km | $\rho_c$, $10^{14}$ g/cm$^3$ | $M_{core}$, $M_\odot$ | $R_{core}$, km | $R$, km |
| 0.02 | 2273.61 | 22759.2 | 4.02716 | 0.01838 | 2.800 | 3086.8 |
| 0.03 | 4043.80 | 23117.1 | 4.02710 | 0.01836 | 2.799 | 7813.2 |
| 0.04 | 14721.21 | 20004.3 | 4.02707 | 0.01835 | 2.798 | 9384.2 |
| 0.05 | 26257.51 | 18109.4 | 4.02706 | 0.01834 | 2.797 | 10094.0 |
| 0.1 | 99958.6 | 14155.6 | 4.02698 | 0.01831 | 2.796 | 10799.6 |
| 0.15 | 194071.7 | 12849.9 | 4.02690 | 0.01828 | 2.795 | 10462.5 |
| 0.25 | 550170.1 | 10890.1 | 4.02672 | 0.01821 | 2.791 | 9509.9 |
| 0.5 | 3.44E+06 | 8000.2 | 4.02597 | 0.01791 | 2.776 | 7291.2 |
| 0.8 | 3.37E+07 | 5154.9 | 4.02346 | 0.01692 | 2.724 | 4506.9 |
| 0.96 | 1.65E+08 | 3565.6 | 4.01591 | 0.01405 | 2.561 | 2347.0 |

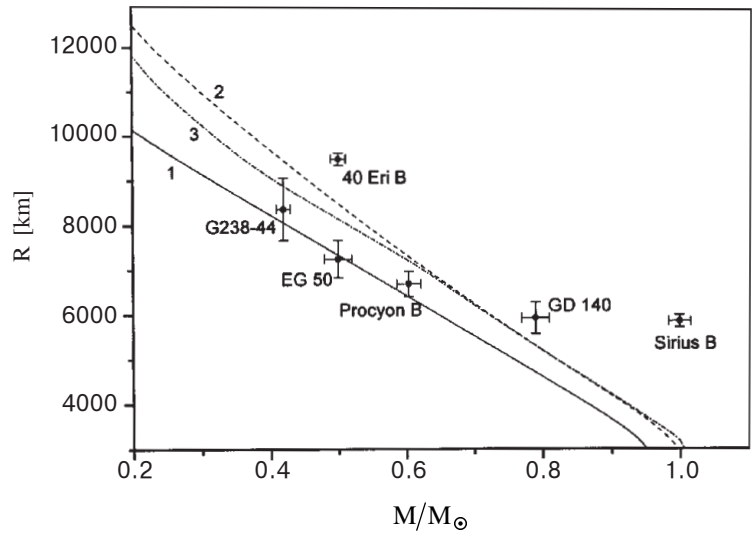

Fig. 5. Radius $R$ as a function of total mass $M$ for strange dwarfs (1, $\rho_{cr} = 4.3 \cdot 10^{11}$ g/cm$^3$, smooth curve; 2, $\rho_{cr}=10^9$ g/cm$^3$, dashed curve) for model 2 and for white dwarfs (3, dashed curve dot-dashed). The dots (·) denote observational data on the mass and radius for six white dwarfs.

Recently, as a result of improved exoatmospheric astronomy, the accuracy with which such integral parameters of stars as their mass and radius can be measured has increased by an order of magnitude. New observational data on white dwarfs obtained through the HIPPARCOS project have made it possible to refine the measured mass and radius for twenty of these objects [24]. Figure 5 shows the radius as a function of mass for strange and white dwarfs. Observational data on the six white dwarfs with the smallest ranges of mass and radius are indicated on this figure, which shows that EG 50, G 238-44, and Procyon B are the most likely candidate strange dwarfs. Refined observational data and comparison of these with theoretical calculations will, in the future, establish whether strange dwarfs can exist.

This work was supported by the Armenian National Science and Education Foundation (ANSEF Grant No. PS 140) and conducted in the framework of topic #0842 financed by the Ministry of Education and Science of the Armenian Republic.